\documentstyle[twocolumn,floats,aps,prl,epsfig]{revtex}
\tighten
\begin{document}
\draft

\title{Finite-Size Effects on Nucleation in a 
First-Order Phase Transition}

\author{
Eduardo S. Fraga$^{1}$ 
and Raju Venugopalan$^{2}$
} 

\address{
$^{1}$Instituto de F\'\i sica, Universidade Federal do Rio de 
Janeiro \\
 C.P. 68528, Rio de Janeiro, RJ 21941-972, Brazil\\
$^{2}$Department of Physics and RIKEN-BNL Research Center,\\
Brookhaven National Laboratory, Upton, NY 11973-5000, USA
}

\date{\today}
\maketitle   


\begin{abstract}
We discuss finite-size effects on homogeneous nucleation in first-order 
phase transitions. We study their implications for  
cosmological phase transitions and to the hadronization of a quark-gluon 
plasma generated in high-energy heavy ion collisions. Very general 
arguments allow us to show that the finite size of the early universe 
has virtually no relevance in the process of nucleation and in the 
growth of cosmological bubbles during the primordial quark-hadron 
and the electroweak phase transitions. In the case of high-energy heavy 
ion collisions, finite-size effects play an inportant role 
in the late-stage growth of hadronic bubbles.
\end{abstract}


\section{Introduction}

First-order phase transitions and the kinetic phenomena 
associated with the process of phase conversion through 
the nucleation of bubbles or spinodal decomposition are 
present in almost all realms of physics \cite{reviews}. Usually, 
the interesting physical quantities are the time scales 
related to the transient regime, since they determine its relative 
importance. However, 
one cannot often disentangle those scales from other scales of the problem, such as 
the different length scales and, in particular, the 
finite size of the system under consideration.

Despite the success of finite-size scaling in the study 
of equilibrium critical phenomena \cite{goldenfeld}, 
a systematic study of finite size effects has rarely been conducted in the case 
of metastable decays and other nonequilibrium processes 
(see Ref. \cite{rikvold} and references therein). In particular, 
a careful analysis of possible modifications of the usual 
picture of nucleation in a thermally-driven first-order phase transition 
due to the finite size of the system seems to be lacking.

In this paper, we discuss finite size effects on the dynamics of 
homogeneous nucleation in a first-order temperature-driven transition. 
In particular, we consider the case of cosmological phase transitions 
in the early universe \cite{cosmo}, and that of a quark-gluon plasma 
(QGP) decay into hadronic matter in a high-energy 
heavy ion collision \cite{rhic,QGPnucleation}. 
The former might provide sensible 
mechanisms to explain the baryon number asymmetry in the universe 
and primordial nucleosynthesis \cite{baryogenesis,nucleosynthesis}, 
whereas the latter is expected to be observed  
\cite{QM2001} at BNL Relativistic 
Heavy Ion Collider (RHIC). The length and time scales involved in each of 
these cases differ by several orders of magnitude. We will argue that,
despite statements to the contrary in the literature, the finite size of the early 
universe plays no important role in the process of homogeneous 
nucleation during the quark-hadron and the electroweak cosmological 
phase transitions. For an expanding QGP plasma, finite size  
effects might prove to be relevant if the plasma is formed in relativistic 
heavy ion collisions.

In the usual description of homogeneous nucleation \cite{reviews}, 
there are two ways in which the finite size of the system 
can affect the formation and evolution of bubbles and, consequently, 
the dynamics of phase conversion. Firstly, one has to consider 
the effects on the nucleation rate and the early stage growth of the 
bubbles. As will be shown below, this correction comes about through 
an intrinsic uncertainty in the determination of the supercooling 
undergone by the system. For the cases considered here, it brings 
only minor modifications to a description which assumes an infinite system. 
The second and, in general, most important finite-size effect is its 
influence on the domain coarsening process, or late-stage growth 
of the bubbles. The relevant length scales here are given by the
typical size of the system, the radius of the critical bubble and 
the correlation length. One has to compare these scales, in each 
case of interest, to measure the relevance of possible 
corrections due to the finite size of the system.

The arguments presented in this paper 
are based solely on analytic considerations and simple estimates. 
To obtain precise results, one should perform numerical simulations. 
Lattice methods have been successfully applied to the study of homogeneous 
nucleation in different contexts~\cite{rikvold,sourendu,lattice}. 
A systematic lattice investigation of bubble nucleation and 
spinodal decomposition, including finite-size effects, 
in the case of high-energy heavy ion collisions is certainly important and 
will be reported in a future publication~\cite{next}.

The paper is organized as follows. In Section II, we present a short review 
of relevant aspects of homogeneous nucleation that are important for 
our analysis. In section III, we discuss the effects of the finite 
size of the system in general, and apply the results to the cases 
mentioned above. Section IV contains our conclusions and outlook.


\section{Homogeneous nucleation in a first-order phase transition}
\label{nucleation}

In a continuum description of a first-order phase transition, it is 
convenient to make use of a coarse-grained free energy, $F$, 
which is expected to have  the familiar Landau-Ginzburg form with 
temperature-dependent coefficients \cite{reviews}. 
(In the case of QCD, such a free energy can be obtained, for instance, 
from the one-loop effective potential of a linear sigma model coupled to 
quarks \cite{Scavenius:1999zc,Scavenius:2001bb}.) 
For temperatures between the critical temperature, 
$T_c$, and a temperature that characterizes the spinodal 
region, $T_{sp}$,  the minimum corresponding to the symmetric phase 
is metastable, and gradually disappears as the system approaches 
$T_{sp}$, where the symmetric phase becomes unstable. In other words, 
at $T=T_{sp}$, the barrier that separates the metastable symmetric phase 
minimum from the true vacuum, within this temperature range, vanishes. 
The driving mechanism for the phase transition above the spinodal 
temperature is the nucleation of large localized domains, or bubbles, 
of the true vacuum phase, 
inside the so-called false vacuum or metastable phase, via 
thermal activation. Below $T_{sp}$, one has the phenomenon of 
spinodal decomposition, where the transition is driven by 
long-wavelength small fluctuations \cite{reviews}. An 
expanding system will probe these 
two regions at a pace given by its rate of expansion. The 
investigation of the time scale for thermal nucleation relative to that 
for the expansion, to evaluate the role played by nucleation 
in the phase conversion, is therefore an interesting issue (see, for instance, Ref. 
\cite{Scavenius:2001bb}). However, for a more complete and consistent 
analysis, one also has to incorporate the effects due to the finite size 
of the system under investigation.

The nucleation rate, namely, the probability per unit volume, per unit time, 
to form a critical bubble, 
can be expressed as $\Gamma={\cal P}~ e^{-F_b/T}$, 
where $F_b$ is the free energy of a critical bubble. The
prefactor ${\cal P}$, discussed further below, measures statistical and 
dynamical fluctuations
about  the saddle point of the Euclidean action in functional space. The 
critical bubble, with radius $R=R_c$, 
is a field configuration given by a radially symmetric, static solution 
of the Euler-Lagrange field 
equations corresponding to an 
exact balance between competing volume and surface energy contributions. 
The critical bubble is unstable with respect to 
small changes of its radius; for subcritical bubbles, 
$R < R_c$, the surface energy dominates, and the bubble shrinks 
and vanishes, while for supercritical bubbles, $R>R_c$, the 
volume energy dominates, and the bubble grows driving the decay process.

It is convenient to write the prefactor ${\cal P}$ as a product of the 
bubble's growth rate and a factor proportional 
to the ratio of the determinant of the fluctuation operator around the bubble 
configuration relative to that around the homogeneous metastable state 
\cite{langer}. For the relativistic case, in the thin-wall limit, 
we have \cite{prefactor}:
\begin{equation}
{\cal P}=\frac{16}{3\pi}\left(\frac{\sigma}{3T}\right)^{3/2}
\frac{\sigma\eta R_c}{\xi^4 (\Delta\omega)^2}\quad.
\end{equation}
Here, $\eta$ and $\xi$ are respectively the shear viscosity and the correlation length 
in the symmetric phase and $\Delta\omega$ is the enthalpy 
density difference between the two phases. $\sigma$ is the surface tension 
of the interface, which is related to the critical radius and the difference 
in pressure between the mestastable phase and the true vacuum, $\Delta p$, as 
$R_c=2\sigma/\Delta p$. The thin-wall limit is a good approximation in 
the limit of small supercooling, $\theta=1-T/T_c\ll 1$. In this limit, the 
decay rate can be written as
\begin{equation}
\Gamma={\cal P}~exp\left( -\frac{4\pi\sigma}{3T}R_c^2\right)=
{\cal P}~exp\left(-\frac{16\pi}{3}\frac{\sigma^3}{T_c \ell^2 \theta^2}\right)
\quad ,
\end{equation}
where $\ell$ is the latent heat density. For 
increasing values of the supercooling, $\theta$, the argument of the 
exponential decreases and the decay rate increases. Equivalently, the 
critical radius becomes smaller, so that it is easier to nucleate 
supercritical bubbles. The only way in which the finite size of the 
system can affect the decay rate is through its influence on the 
supercooling. (Of course, one has to check whether the size of the 
system allows for the presence of at least one critical bubble, 
including the wall that separates the two phases, or not.) 


\section{Finite-size effects}

\subsection{Coarse-grained free energy}

Before considering the effects of the finite size 
of the system on the nucleation process, let us move one step backwards to 
examine a point that will be relevant for the following, namely, the definition 
of metastable states 
in equilibrium statistical mechanics. 
Since the complete partition function is dominated by  
configurations that minimize the free energy and correspond to equilibrium states, 
it does not allow for the 
existence of metastable branches in the thermodynamic limit. One 
solution to this difficulty consists in constraining the phase space and 
computing {\it restricted partition functions}. In order to 
define an optimal cut, one has to resort to physical arguments \cite{reviews}.

Such a procedure can be implemented as follows \cite{langerCG}. 
The system of size $L$ is divided into cells of linear size $\lambda_{cg}$ 
centered at positions $\vec{x}$. The length scale which defines 
the coarse-graining of the system, $\lambda_{cg}$, should be 
appreciably larger than the underlying lattice spacing $a$. 
Moreover, within each cell, the relevant order parameter, $\phi$, should 
vary smoothly in space and the equilibration time of the system 
should be much faster than the processes under consideration. This allows us 
to adopt a continuum order parameter, $\phi(\vec{x},t)$, defined as 
an average over each cell. 

However, by construction, the coarse-grained free energy will depend on the 
scale $\lambda_{cg}$ 
and, as mentioned previously, one has to appropriately constrain the phase 
space to allow for metastable states. 
If $\lambda_{cg}$ is small enough as compared to the correlation 
length, $\xi$, phase separation within a cell cannot occur and one can 
define a coarse-grained free energy, $F$, such as the one we 
considered in the discussion of Section \ref{nucleation}. 
The condition 
$\lambda_{cg}<<\xi$ therefore represents the physically motivated restriction 
to the partition function that allows for a well-defined 
coarse-grained description of metastable states. As a corollary, 
increasing the value of the coarse-graining length relative to the correlation 
length would incorporate 
more and more fluctuation modes, which were excluded by this 
ultraviolet cut-off. The free energy $F$ would then flatten out completely in the 
limit $\lambda_{cg}/\xi\to\infty$, thereby approaching the equilibrium 
description with no metastable branch~\cite{largeL}. 
For a system to be characterized by a well-defined Landau-Ginzburg coarse-grained 
free energy one therefore requires a clean separation of 
scales: $a << \lambda_{cg} << \xi << L$.

\subsection{Nucleation rate}

The only effect that might be relevant to the nucleation rate, 
$\Gamma$, is the rounding of singularities, since they will 
affect the degree of supercooling. The relevant physical quantity 
can be calculated in a finite-size 
scaling framework \cite{binder}, resulting in the following expression:
\begin{equation}
\frac{\Delta T_{round}}{T_c}\approx \frac{2T_c}{\ell L^d}\quad ,
\label{round}
\end{equation}
where $L$ is the typical length scale of the system and $d$ is the 
number of dimensions. The quantity $\Delta T_{round}$ is a measure of 
the smoothening of singularities due to the finite size of the system 
\cite{greiner}. 
It corresponds to the (now non-vanishing) width in temperature of the 
region where the energy density of the system suffers an abrupt jump, around 
$T=T_c$ \cite{shift}. This phenomenon enters the decay rate 
as a minimum value for $\theta$ which reflects the uncertainity in 
the supercooling 
due to the finite size of the system. 
This conclusion should be contrasted with the arguments presented in
Ref. \cite{gavai}, particularly in the case of cosmological 
phase transitions \cite{discussion}.

For phase transitions in the early universe, $L$ is given by 
the radius of the universe at a given time (or temperature). Going back to 
very early times, $t \sim 10^{-4}~$s and 
$T \sim 10^{12}~$K$~\sim 10^{-1}~$GeV, we can use the elementary particle 
model \cite{weinberg} to obtain an approximate equation of state (ideal gas) 
\begin{equation}
3p\approx \epsilon \approx \frac{\pi^2}{30}N(T)T^4 \quad ,
\end{equation}
where $N(T)=N_{bosons}(T)+(7/8)N_{fermions}(T)$ is the 
total number of degrees of freedom, $\epsilon$ is the energy 
density and $p$ is the pressure. From Einstein's field equations, one can 
derive the relation between time and energy density in the early universe, 
which links the age of the universe and its temperature in the following way
\begin{equation}
t\approx \left( \frac{3}{32\pi G\epsilon}  \right)^{1/2}\approx 
\frac{1}{4\pi}\left( \frac{45}{\pi N(T)}  \right)^{1/2} \frac{M_{Pl}}{T^2}
\quad ,
\end{equation}
where $M_{Pl}$ is the Planck mass and $G$ is Newton's gravitational 
constant. The radius of the universe, as given 
by the particle horizon in a Robertson-Walker spacetime \cite{kolb}, 
$d_h(t)=t/(1-n)$, where the scale factor grows as 
$a(t)\sim t^{2/3(1+w)}=t^n$ and $p=w\epsilon$, has the following form:
\begin{equation}
L_{univ}(T)\approx \frac{1}{4\pi}\left( \frac{1}{1-n}  \right)
\left( \frac{45}{\pi N(T)}  \right)^{1/2} \frac{M_{Pl}}{T^2} \quad .
\end{equation}
For the assumed equation of state, $w=1/3$ and $n=1/2$. If we enter 
temperature in GeV, the typical length scale of the early universe is given, 
in fermi units, by $L_{univ}(T)\approx {\cal A}/T^2\sqrt{N(T)}$, where 
${\cal A}=1.45 \times 10^{18}$.

It is clear that, due to the large factor $\sim 10^{18}$ (coming from 
$M_{Pl}$), $L_{univ}$ will be of importance for supercooling only 
for extremely high values 
of the temperature. Since $N(T)\approx 50$ for the cases of the QCD and the 
electroweak phase transitions, one would need $T\sim 10^8~$GeV to be 
subject to appreciable finite-size effects on the nucleation rate. Therefore, 
since $T_c^{QCD}\sim {\cal O}(10^{-1})~$GeV and 
$T_c^{EW}\sim {\cal O}(10^{2})~$GeV,
these effects are negligible in such cases. They might prove to be relevant 
in the case of GUT's transitions, where the critical temperature will be 
much higher.

For the first order transition of a quark-gluon plasma into a hadronic gas 
in a high-energy heavy ion collision, one can easily estimate the 
role of finite-size effects on the nucleation rate. The parameters entering 
Eq. (\ref{round}) are approximately given by $T_c\approx 150~$ MeV 
\cite{Karsch:2001vs}, 
$\ell\approx 4B \approx 4 \times (150~$ MeV$)^4$, where $B$ is the bag 
constant, and $L \approx 10~$ fm. The minimal amount of supercooling 
undergone by the plasma is then negligible, of the order of $10^{-3}$. 
(For more conservative, smaller values of $L$, the minimum value 
of $\theta$ is still very small.)  
The system can in principle probe the entire domain in $\theta$, and there 
are no constraints on homogeneous nucleation from this side. The question 
of how fast the system probes the nucleation region, before reaching the 
spinodal regime, as compared to the nucleation rate 
was discussed in 
Ref. \cite{Scavenius:2001bb,bravina,Shukla:2001xv}. 
Recently, it has been speculated that the RHIC data suggest 
an {\it explosive} hadron production due to a rapid variation of the 
effective potential for QCD close to $T_c$. The theoretical reasoning is 
based on the results of the Polyakov Loop Model \cite{polyakov} for 
the deconfining phase transition, which lead to a very fast spinodal 
decomposition regime \cite{explosive}.


\subsection{Late-stage growth}

After the nucleation of a given supercritical bubble, it will grow with a 
certain velocity. The set of all supercritical bubbles created 
integrated over time will eventually drive the complete phase conversion in a 
finite system. The scales that determine the importance of 
finite-size effects are the typical linear size of the system, 
the radius of the critical bubble and the correlation length. For the 
reasons discussed in the last subsection, the case of cosmological phase 
transitions is to an excellent approximation free from finite-size 
corrections. The enormous numerical value of the Planck mass washes out 
every other scale. Therefore, throughout this subsection we will address the 
quark-gluon--hadron phase transition in heavy ion collisions.

For definiteness, let us assume our system is characterized by 
a coarse-grained free energy of the form
\begin{equation}
F(\phi,T)=\int ~d^d x \left[ \frac{1}{2}(\nabla\phi)^2 + U(\phi,T)  \right]
\quad,
\end{equation}
where $U(\phi,T)$ is some Landau-Ginzburg potential whose coefficients 
depend on the temperature, and $\phi(\vec{x},t)$ is a scalar field 
\cite{orderparameter}. 
For the cases to be considered in this paper, 
the order parameter, $\phi$, is {\it not} a conserved quantity, and 
its evolution is given by the time-dependent Landau-Ginzburg equation 
\cite{reviews}
\begin{equation}
\frac{\partial\phi}{\partial t}=-\gamma \frac{\delta F}{\delta\phi}=
\gamma \left[ \nabla^2\phi - U'(\phi,T) \right]\quad ,
\label{reaction}
\end{equation}
where $\gamma$ is the response coefficient which defines a time scale for 
the system. Equation (\ref{reaction}) is a standard reaction-diffusion 
equation, and describes the approach to equilibrium \cite{langevin}.

If $U(\phi,T)$ is such that it allows for the existence of bubble solutions 
(taken to be spherical for simplicity), then supercritical (subcritical) 
bubbles expand (shrink), in the thin-wall limit, with the following 
velocity:
\begin{equation}
\frac{dR}{dt}=\gamma (d-1) \left[ \frac{1}{R_c}-\frac{1}{R(t)}  \right]
\quad ,
\label{allen-cahn}
\end{equation}
where $R_c=(d-1)\sigma/\Delta F$ and 
$\Delta F$ is the difference in free energy 
between the two phases. Equation (\ref{allen-cahn}) is an example of the 
Allen-Cahn equation \cite{reviews}, 
which relates the velocity of a domain wall to the local 
curvature. The response coefficient, $\gamma$, can be related to some 
characteristic collision time as will be done later.

The description of the late-stage domain coarsening is given by 
the Kolmogorov-Avrami theory \cite{reviews}, which contains the 
following assumptions: (i) bubbles grow without substantial deformation and 
are uncorrelated; (ii) the nucleation rate is a constant; (iii) the 
bubble growth velocity is constant, $v=dR/dt=\gamma (d-1)/R_c$ (limit 
$R\to\infty$). The quantity which is usually computed is the volume 
fraction of the stable phase. 

However, one can measure the importance of finite-size effects for 
the case of heavy-ion collisions by comparing, for instance, the 
asymptotic growth velocity ($R>>R_c$) 
for nucleated hadronic bubbles to the 
expansion velocity of the plasma. In the Bjorken picture, one assumes 
that the central rapidity region exhibits longitudinal expansion, so 
that $z(t)=v_z t$, where $v_z$ is the collective fluid velocity. 
Conservation of total entropy leads to adiabatic expansion and the 
following cooling behavior: $(t/t_i)=(T_i/T)^3$. The typical 
length scale of the expanding system is then 
\begin{equation}
L(T)\approx (v_z t_c)\left(\frac{T_c}{T}\right)^3=
L_0 \left(\frac{T_c}{T}\right)^3 \quad ,
\label{Lqgp}
\end{equation}
where $L_0\equiv L(T_c)$ is the initial linear scale of the 
system for the nucleation process which starts at $T\leq T_c$.

The relation between time and temperature provided by the 
cooling law that emerges from the Bjorken picture suggests 
the comparison between the following ``velocities'':
\begin{equation}
v_b\equiv \frac{dR}{dT}=
-\left(\frac{3b\ell L_0}{2v_z\sigma T_c^2}\right)
\left(\frac{T_c}{T}\right)^5 \left( 1-\frac{T}{T_c} \right)\quad , 
\end{equation}
the asymptotic bubble growth ``velocity'', and the plasma 
expansion ``velocity''
\begin{equation}
v_L\equiv \frac{dL}{dT}=
-\frac{3L_0}{T_c}\left(\frac{T_c}{T}\right)^4\quad . 
\end{equation}

The quantity $b$ is a number of order one to first approximation, 
and comes about in the estimate of the phenomenological 
response coefficient $\gamma (T)\approx b/2T$ 
(see Ref. \cite{kajantie} for details). 

Using the numerical values adopted in the previous section and 
$\sigma/T_c^3\sim 0.1$, we obtain
\begin{equation}
\frac{v_b}{v_L}\approx \frac{20}{v_z}
\left( \frac{T_c}{T} -1\right)\quad .
\end{equation}
One thus observes that the bubble growth velocity becomes 
larger than the expansion velocity for a supercooling of 
order $\theta\approx v_z/20 \leq 5\%$. A simple estimate 
points to a critical radius larger than $1~$fm at 
such values of supercooling (see also \cite{Scavenius:2001bb}). 
Therefore, finite-size effects appear to be an important 
ingredient in the phase conversion process right from the 
start in the case of high-energy heavy-ion collisions.


\section{Conclusions and outlook}

We have studied the effects of the finite size of the 
system on the nucleation of bubbles in thermally-driven 
first-order phase transition. The physical situations 
we considered were those of the cosmological transitions 
in the early universe and the hadronization of a 
quark-gluon plasma after a high-energy heavy ion 
collision. We have shown, on very general grounds, that 
during the quark-hadron and the electroweak cosmological 
transitions the universe is large enough so that one 
is safe to approximate it by an infinite system. 
Corrections due to its finite size are negligible. For the 
case of heavy ions, although the influence on the 
nucleation rate is also small, effects related to 
domain coarsening and late-stage growth are  
important and should be taken into account in a detailed 
study.

In order to address the question of hadronization 
after a first-order transition, one should 
then perform finite-size real-time lattice simulations. 
One would thereby avoid the drawbacks implied by  
analytical approximations, such as the thin-wall 
hypothesis. Moreover, one would be able to control the 
behavior of domains and study, for instance, scaling 
properties \cite{next}. It is also interesting to 
study the hydrodynamics of nuclear matter at chiral 
limit as a phenomenological description of the chiral 
transition in an expanding quark-gluon plasma 
\cite{son,ove,adrian,AFK}. Results in this direction 
will be presented elsewhere.

 

We thank A. Dumitru, R. Gavai, S. Gupta 
and A. Krasnitz for fruitful discussions. E.S.F. thanks 
the members of LPT (Orsay), where part of this work has been 
done, for their kind hospitality. 
E.S.F. is partially supported by FAPERJ and FUJB/UFRJ. 
R.V. is supported by the U.S. Department of Energy under Contract
No. DE-AC02-98CH10886 and by the RIKEN-BNL Research Center at BNL. 



\begin{thebibliography}{10}
\bibitem{largeL}Note that this unphysical limit should not be confused with the 
thermodynamic limit of $L\rightarrow \infty$.
\bibitem{shift} There is also a shift in the value of the critical 
temperature as a result of the finite size of the system \cite{binder}. 
Here, $T_c$ denotes the corrected value.
\bibitem{orderparameter} It is implicit in our discussion that we restrict 
ourselves to the case in which the order parameter is a scalar ($N=1$ in the 
language of critical phenomena), and that we assume that there is 
a maximum of two coexisting thermodynamic phases.
\bibitem{langevin} In order to include the effect of fluctuations, one should 
add a Langevin noise term in (\ref{reaction}).
\end{thebibliography}
\end{document}